\documentclass[9pt,a4paper]{article}
\usepackage[scaled]{helvet}
\usepackage[applemac]{inputenc}
\usepackage{geometry}
\usepackage{url}
\usepackage{siunitx}
\usepackage[super,numbers,sort&compress]{natbib}
\makeatletter
\renewcommand\@biblabel[1]{#1.}
\makeatother
\usepackage{setspace} 
\usepackage{color}
\usepackage{amsmath}
\usepackage{amsfonts}
\usepackage{amssymb}
\usepackage{graphicx}
\usepackage{multicol}
\usepackage{lineno}
\usepackage{caption}
\include{00README.XXX}

\newcommand{\ket}[1]{\left| #1 \right\rangle}

\DeclareCaptionLabelSeparator{bar}{ $\vert $  }

\makeatletter
\renewcommand{\maketitle}{\bgroup\setlength{\parindent}{0pt}
\begin{flushleft}
  \textbf{\Huge \@title} \\[0.5 cm]

  \@author
\end{flushleft}\egroup
}

\author{Benedict Seiferle$^1$, Lars von der Wense$^1$, Pavlo V. Bilous$^{2}$, Ines Amersdorffer$^1$, Christoph Lemell$^{3}$, Florian Libisch$^{3}$, Simon Stellmer$^{4}$, 
Thorsten Schumm$^{5}$, 
Christoph E. D\"ullmann$^{6,7,8}$, Adriana P\'alffy$^{2}$ \&  Peter G. Thirolf$^1$
\footnote{ 
\noindent$^{1}$Ludwig-Maximilians-University Munich, 85748 Garching, Germany. 
$^{2}$Max-Planck-Institut f\"ur Kernphysik, Heidelberg, 69117 Heidelberg, Germany.  
$^{3}$ Inst. for Theoretical Physics, TU Wien, 1040 Vienna, Austria.
$^{4}$ University of Bonn, 53105 Bonn, Germany. 
$^{5}$ Atominstitut, TU Wien, 1020 Vienna, Austria. 
$^{6}$ GSI Helmholtzzentrum f\"ur Schwerionenforschung GmbH, 64291 Darmstadt, Germany. 
$^{7}$ Helmholtz Institute Mainz, 55099 Mainz, Germany. 
$^{8}$ Johannes Gutenberg University, 55099 Mainz, Germany.} }

\title{Energy of the $^{\text{\large 229}}$Th nuclear clock transition}
\begin{document}
\maketitle
\noindent

\begin{spacing}{1.5}

\noindent\textbf{%
The first nuclear excited state of $^{229}$Th offers the unique opportunity for laser-based optical control of a nucleus\cite{Kroger_Reich,Lars}.
Its exceptional properties allow for the development of a nuclear optical clock\cite{Peik1} which offers a complementary technology and is expected to outperform current electronic-shell based atomic clocks \cite{Campbell2}.
The development of a nuclear clock was so far impeded by an imprecise knowledge of the energy of the $^{229}$Th nuclear excited state.
In this letter we report a direct excitation energy measurement of this elusive state and constrain this to 8.28$\pm$0.17 eV.  
The energy is determined by spectroscopy of the internal conversion electrons emitted in-flight during the decay of the excited nucleus in neutral $^{229}$Th atoms. 
The nuclear excitation energy is measured via the valence electronic shell, thereby merging the fields of nuclear- and atomic physics to advance precision metrology.
The transition energy between ground and excited state corresponds to a wavelength of 149.7$\pm$3.1 nm. 
These findings set the starting point for high-resolution nuclear laser spectroscopy and thus the development of a nuclear optical clock of unprecedented accuracy.
A nuclear clock is expected to have a large variety of applications, ranging from relativistic geodesy\cite{Mehlstaubler} over dark matter research\cite{Derevianko} to the observation of potential temporal variation of fundamental constants\cite{Flambaum1}.\newline \\}
\noindent

%
%

The first excited isomeric state of $^{229}$Th, denoted by $^{229\text{m}}$Th, has the lowest excitation energy of all known nuclear states. 
While typical transition energies in nuclear physics range from several keV to MeV, the excitation energy of $^{229\text{m}}$Th is in the eV region\cite{Reich, Helmer, Beck1, Beck2}.
$^{229\text{m}}$Th possesses a long radiative lifetime of expectedly up to 10$^4$~s (refs. \citenum{Tkalya1, Minkov}) resulting in a narrow relative linewidth of $\Delta E/E\approx10^{-20}$.  \\
These properties of the excited state make $^{229\text{m}}$Th the only candidate for a new type of optical clock that uses a nuclear transition instead of an electronic transition for time measurement\cite{Peik1, Campbell2}. 
While posing a complementary technology to existing atomic clocks, a single-ion nuclear optical clock is expected to achieve a systematic frequency uncertainty of 1.5$\times10^{-19}$ (ref. \citenum{Campbell2}), thereby reaching and even surpassing existing atomic clock technology.
Moreover, the possibility to dope $^{229}$Th nuclei into a vacuum ultra-violet (VUV) transparent host crystal could allow for the operation of a solid-state optical clock, which profits from the high density of nuclei that can be addressed\cite{Rellergert}.
During the past years $^{229\text{m}}$Th has been subject to vivid research. This includes the theoretical prediction of the isomeric properties \cite{Tkalya1,Minkov, Karpeshin1}, as well as experimental work aiming for further characterisation of the isomer\cite{Burke2,Raeder,Jeet,Yamaguchi,Stellmer2, Borisyuk, Stellmer, Kazakov,Campbell,Bilous1, Lars2}.

\noindent Until today the energy of $^{229\text{m}}$Th has been exclusively inferred from indirect measurements\cite{Kroger_Reich, Reich, Helmer, Beck1, Beck2} probing the gamma emission from higher lying excited states populating the ground and isomeric state.
The latest of these measurements constrained the energy to 7.8$\pm$0.5 eV\cite{Beck2}.
The uncertainty of this result is due to the detectors' energy resolution. 
A direct detection of the isomeric decay in the internal conversion decay channel\cite{Lars} already led to the determination of the isomeric half-life in neutral, surface-bound, $^{229\text{m}}$Th atoms\cite{Seiferle}.
Additionally, laser spectroscopic characterisation of the isomeric state has been achieved recently\cite{Thielking}.\\
In this letter we report the first direct energy measurement of the $^{229}$Th nuclear isomer.
We use the internal conversion (IC) decay of the isomeric excited state in a $^{229}$Th atom. 
In this decay channel the nuclear energy is transferred to the electronic shell with ionisation of the atom.
The emitted electron's kinetic energy can be measured, which allows to deduce the energy of the isomer.
This approach offers the advantage that it relies on the atomic structure of the Th atom 
which is on the same energy scale as the isomer's energy. 
Our results are sufficiently precise to develop a laser for the direct excitation of the isomeric state in $^{229}$Th and paves the way for a nuclear clock.
\\[0.3 cm]
%
%
%
%
\begin{figure*}[bt]
\begin{center}
\includegraphics[width = 0.95\textwidth]{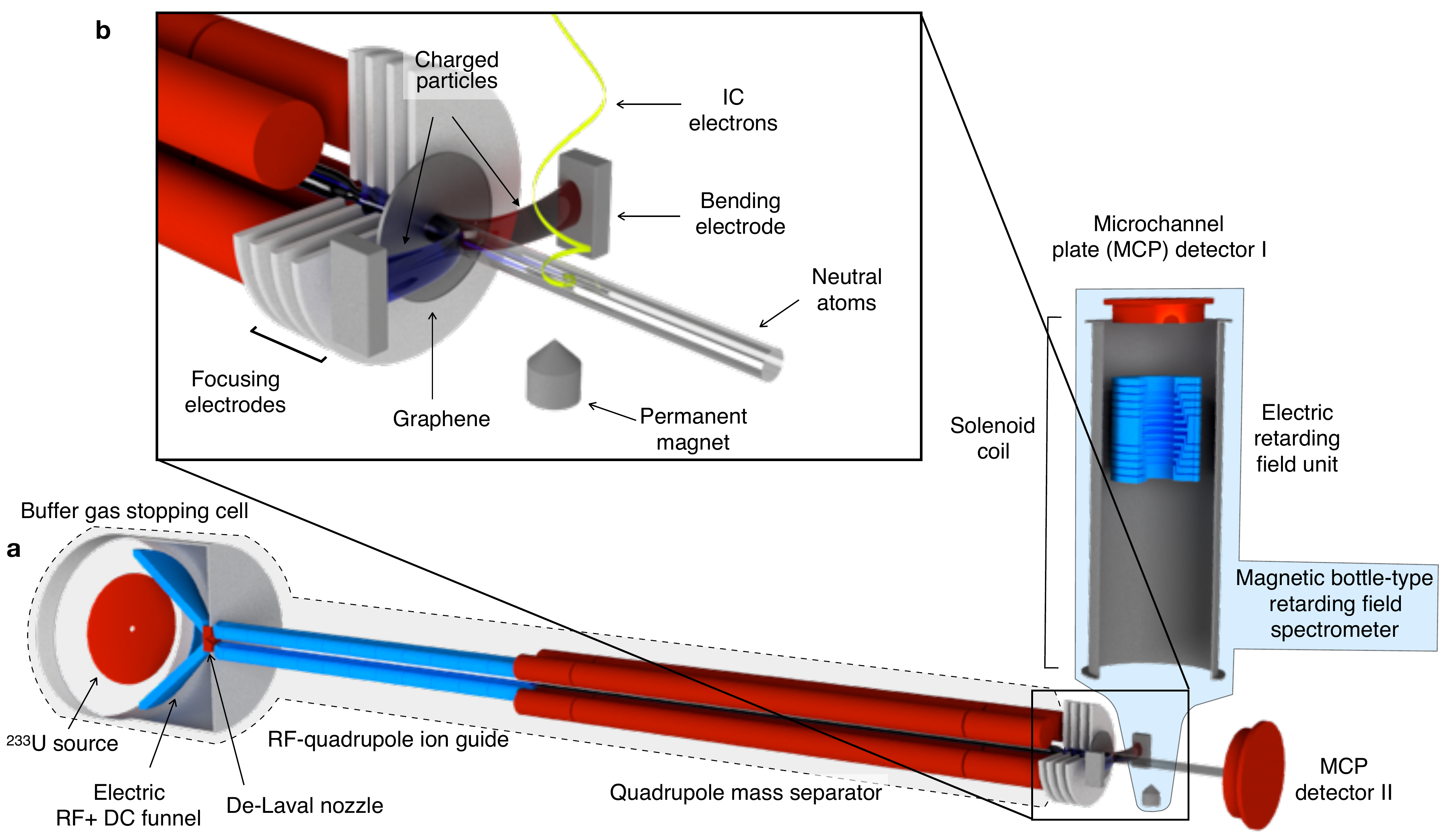}
\end{center}
\caption{\textbf{Schematic of the experimental setup used for the energy determination of $^{229\text{m}}$Th.} The experimental setup can be structured in three parts: the ion extraction (framed with the dashed line), neutralisation (shown in inset \textbf{b}) and electron spectrometer setup (in front of a light blue shaded background). 
A $^{233}$U source is placed in a buffer gas stopping cell and $^{229\text{(m)}}$Th recoil ions are thermalised  and guided with a funnel-shaped ring electrode structure (RF + DC funnel) towards a de-Laval exit nozzle. 
The ions are then injected into a segmented radio frequency quadrupole structure that allows to form ion bunches. 
A quadrupole mass separator guides the ions to focusing electrodes that collimate the ions which are then neutralised in a graphene foil (see inset \textbf{b}). 
They continue their flight as atoms towards the electron spectrometer while decaying via internal conversion (IC). 
Bending electrodes (to which a DC field is applied) are placed between the graphene layer and the spectrometer entrance in order to prevent charged particles from entering the spectrometer. 
IC electrons, which are emitted above a strong permanent magnet, are collected and guided towards a retarding field unit placed in a solenoid coil. 
The electrons' kinetic energy can be analysed by applying a retarding voltage to a grid and counting the electrons that reach the MCP detector I. 
\label{Fig1}}
\end{figure*}
In our experimental setup, $^{229\text{m}}$Th is generated by a 2\% decay branch in the $^{233}$U $\alpha$-decay. 
The experimental setup for thorium ion-beam and -bunch formation is described in refs. \citenum{Lars, Seiferle} and shown in Fig.~\ref{Fig1}a.
Bunches containing 400 $^{229\text{m}}$Th$^{3+}$ ions are generated at a 10 Hz repetition rate.
The ions are guided by four focusing electrodes onto two layers of graphene set to $-$300 V (graphene layers are supported by lacey-carbon on a copper transmission electron microscopy (TEM) grid, 300 Mesh, with 50 \% transmission).
In passing these foils, the ions are neutralised and continue their flight as neutral atoms (Fig.~\ref{Fig1}b).
The extraction and neutralisation is monitored with a multi-channel plate (MCP) detector placed in the central beam axis (MCP detector II in Fig.~\ref{Fig1}a).
Contrary to the long lifetime of the isomeric state in the Th$^{3+}$ ions, the lifetime in neutral thorium is about 10$^9$ times shorter as the IC channel opens up energetically via the availability of more loosely bound valence electrons. 
Therefore, the isomer decays within microseconds\cite{Seiferle} by emitting an electron.
The electron's kinetic energy is measured free from any surface influences with a magnetic bottle-type retarding field electron spectrometer\cite{Yamakita} which is placed 90$^\circ$ off-axis behind the graphene (see Fig.~\ref{Fig1}).
Secondary particles, such as electrons generated as the ions are passing the graphene or ions which were not fully neutralised, are removed by bending electrodes placed between the point of neutralisation and the spectrometer (see Fig.~\ref{Fig1}b).
The detected electrons can be unambiguously attributed to the nuclear decay of $^{229\text{m}}$Th. 
Comparative measurements that were performed under identical conditions with $^{230}$Th, where no such isomer exists, do not show any comparable signal behaviour (see Extended Data Figure \ref{comparison}).
Therefore, $^{229}$Th atoms in the nuclear ground-state, secondary electrons or auto-ionising states populated in the neutralisation process can be safely excluded as signal origin.\\ \noindent
The spectrometer\cite{Seiferle3} consists of a strong permanent magnet which generates an inhomogeneous magnetic field ($\approx$~200 mT in the region above the magnet) and a solenoid coil that generates a weak homogeneous field (typically 2~mT).
Electrons which are emitted in a spherical region of $\approx$1 mm radius above the permanent magnet are collected by the magnetic field and directed towards the solenoid coil.
In this way a collimated electron beam is generated. 
The kinetic energy of the electrons is analysed by retarding fields, applied to a gold grid (electroformed gold mesh, transmission 90 \%) that is surrounded by ring electrodes to ensure a smooth gradient, and terminated by additional gold grids (Ext.~Data Figure \ref{Spec}).
Electrons whose kinetic energies are sufficient to overcome the applied retarding voltage are counted with an MCP detector (MCP detector I in Fig.~\ref{Fig1}a).
This results in an integrated spectrum which is monotonically decreasing with increasing retarding voltage and in which transition lines are visible as edges. 
The spectrometer reaches a FWHM resolution $(\Delta E /E)$ of about 3\% and its performance and calibration was verified regularly during the measurements. 
\noindent The integrated spectrum of IC electrons as measured by the electron spectrometer is shown in Fig. \ref{effectiveBinding}a.
Data was collected for 3 days in a continuous measurement.\\
\begin{figure*}[bth]
\begin{center}
\includegraphics[width = 1\textwidth]{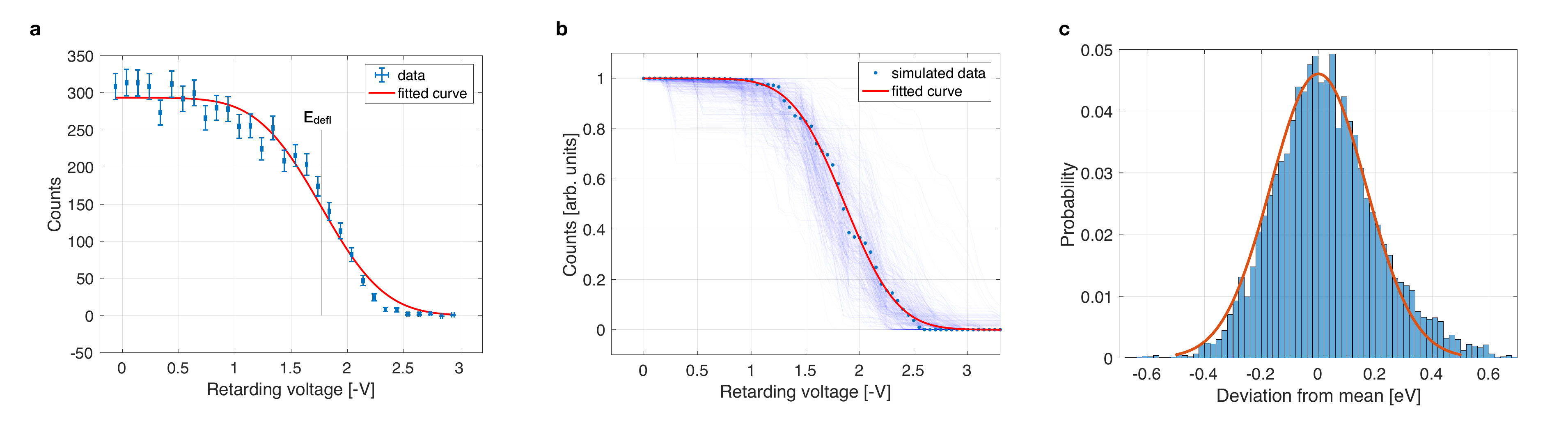}
\caption{\textbf{Measured and predicted internal conversion electron spectra.} \textbf{a} The internal conversion electron spectrum as measured with the magnetic bottle retarding field analyser. Error bars indicate the standard deviation (which is given by the square root of the measured counts). \textbf{b} Visualisation of the expected IC electron energy spectra. Simulated IC spectra (with a simulated isomeric energy $E_{\mbox{\scriptsize sim}} = 8.3$~eV, see text) from 500 different arbitrary selections of five initial excited states (thin blue lines). An exemplary spectrum together with the fitted complementary error function is shown in dark blue. \textbf{c} Scattering of the reference energy $E_0$ around a central value for different initial state distributions. 
20,000 population distributions (from different selections of five initial excited states)  were generated and the resulting reference energies were filled in a histogram. 
The 1$\sigma$ width of the distribution defines the systematic uncertainty of the analysis procedure.
 \label{effectiveBinding}}
\end{center}
\end{figure*}
The neutralisation process of the original Th ion in the graphene layer and after separation from the graphene has to be considered in order to determine the isomeric energy.
Accordingly, all electronic states of Th which have come into resonance with the graphene valence band during and after transit may be occupied\cite{Arnau} (for details on the neutralisation process see Methods).
Thus, IC occurs from excited electronic states:
the kinetic energy of an electron, $E_{\text{kin}}$, which is emitted during IC is connected to the energy of the isomer, $E_I$, via $E_{\text{kin}} = E_I -\text{IP} + E_i - E_f$, with the thorium ionisation potential IP (6.31~eV\cite{Kohler}), $E_i$ the excitation energy of the Th atom undergoing IC and $E_f$ the final electronic state energy of the Th ion generated during the IC process.
Initial-final state pairs are subject to selection rules and each initial state generates a spectrum with a set of lines (see Ext.~Data Fig.~\ref{ExpectEven}).
The sum of these spectra weighted according to the population of the respective initial electronic state results in the measured spectrum.\\
The energy is determined by fitting a complementary error function to the measured data ($f(U) = a (1-\mbox{erf}[ (U-E_{\mbox{\scriptsize defl}})/b)])$, with $U$ being the applied retarding voltage, see Fig. \ref{effectiveBinding}a).
The deflection point of this error function $E_{\text{defl}}$ is determined to $E_{\text{defl}}~=~$1.77$\pm$0.03 ~eV. 
The deflection point shifts according to the isomeric energy, to which it can be related by adding a reference energy $E_0$ such that $E_I = E_{\text{defl}} + E_0$.
The error-function fit is applied to simulated spectra to predict $E_0$.

As the statistical distribution of initial electronic excited states is unknown, we quantify its influence by simulations determining the systematic error. 
Assuming random distributions of initial electronic states, we find that $E_0$ scatters around a central value of $E_0=6.51$ eV and follows a Gaussian distribution (Fig. 2c).  
The width of this Gaussian gets narrower (and thus the uncertainty of the approach improves) with an increasing number of random initial states that make a contribution to the simulation. 
As clear edges resulting from individual lines from a single or only very few
initial electronic states are not observed in our experiment, we
conclude that IC electrons from at least 5 initial states contribute to
the measured spectrum. 
Therefore, to find an upper bound for the
systematic error, we set the number of initial states in the
simulation to $N=5$. 
A collection of simulated spectra is shown in
Fig. 2b (thin lines), the distribution of deviations from $E_0$ in
Fig. 2c.
The uncertainty of 0.16~eV is given by the standard deviation of the normal distribution shown in Fig. \ref{effectiveBinding}c and serves as a conservative upper bound.
Together with the central value $E_0$ this translates to an isomeric energy of $E_I = $8.28$\pm$0.03$_{\mbox{\scriptsize ~stat.}} \pm 0.16_{\mbox{\scriptsize ~syst.}}$ eV.
In total the uncertainties sum up quadratically to give
\begin{equation}
E_I = 8.28 \pm 0.17 \mbox{ eV.}
\end{equation}
Density functional theory calculations indicate that a subset of excited states is populated more dominantly in the neutralisation process.
This would shift the energy to the lower end of the error bar (see Methods).\\
\noindent This measurement represents the first direct energy determination of the lowest nuclear excited state in $^{229}$Th.
The energy corresponds to a wavelength of 149.7$\pm$3.1 nm that is required for a direct optical excitation of the isomeric state.
The wavelength lies in the transparency window of commonly used VUV-transparent crystals (\textit{e.g.} MgF$_2$ or CaF$_2$).
The measured energy value agrees with the former value of 7.8$\pm$0.5 eV\cite{Beck2} within its 1$\sigma$ statistical uncertainty.
Based on our findings the laser technology required for precision laser spectroscopy can be determined.
The wavelength can be reached by high-harmonic generation (\textit{e.g.} the 7$^\text{th}$ harmonic of an Yb-doped fibre laser).
Our improved precision reduces the time required for laser-based scanning in search for the nuclear excitation, \textit{e.g.}, to less than one day using the scheme proposed in ref. \citenum{Lars2}.
This paves the way for nuclear precision spectroscopy and the development of a nuclear optical clock that is expected to have major implications for future frequency metrology.

\begin{small}
\bibliographystyle{nature}

\noindent{\bf Acknowledgements} We acknowledge fruitful discussions with D. Habs, J. Weitenberg, M. Laatiaoui, A. Ulrich, W. Pla{\ss} and colleagues, J. Crespo, C. Weber as well as K. Eberhardt, C. Mokry, J. Runke and N. Trautmann for producing the $^{233}$U and $^{234}$U sources. This work was supported by DFG (Th956/3-2), by the LMU Chair of Medical Physics via the Maier-Leibnitz Laboratory Garching and by the European Union's Horizon 2020 research and innovation programme under grant agreement No 664732 ``nuClock". \\[0.2cm]

\end{small}
%
%
%
%
%
%
%
%
%
%
%
%
%
%
%
\newpage
\noindent\textbf{\Large Methods}\\
\captionsetup[figure]{labelfont={bf},labelformat={default},labelsep=bar,name={Extended Data Figure}}
\captionsetup[table]{labelfont={bf},labelformat={default},labelsep=bar,name={Extended Data Table}}
\setcounter{figure}{0}

\begin{figure*}[ht]
\begin{center}
\includegraphics[scale = 0.45]{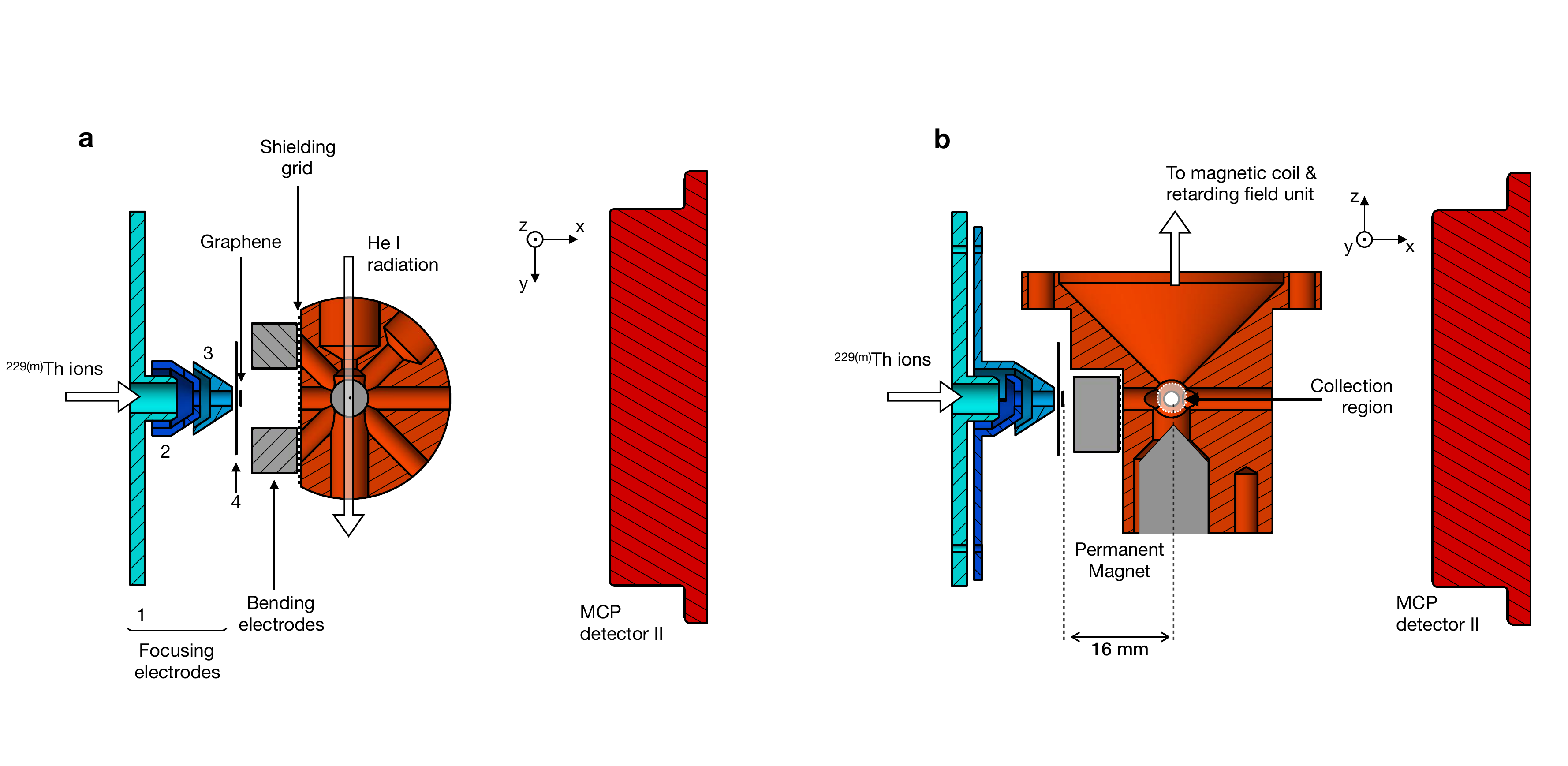}
\end{center}
\caption{Sectional view of the focusing and bending electrodes and the collection region.
Panel \textbf{a} shows the top view, while a lateral view is shown in panel \textbf{b}. Ions are focused by focusing electrodes (labeled with 1,2,3 and 4) onto graphene.
A large amount of the ions is neutralised.
Secondary electrons as well as ions which were not neutralised are deflected by bending electrodes and cannot enter the spectrometer collection region.
A grid which is placed at the entrance of the collection region is used to shield electric RF- and DC fields.
The collection region is accessible via several ports in order to allow for a calibration.
For calibration a gas discharge lamp is used and photoelectrons are generated along a line indicated by the arrow labeled with He I radiation. 
An MCP detector is used to monitor the ion/atom extraction. \label{Triode}}
\end{figure*}

\begin{figure*}[ht]
\includegraphics[scale = 0.4]{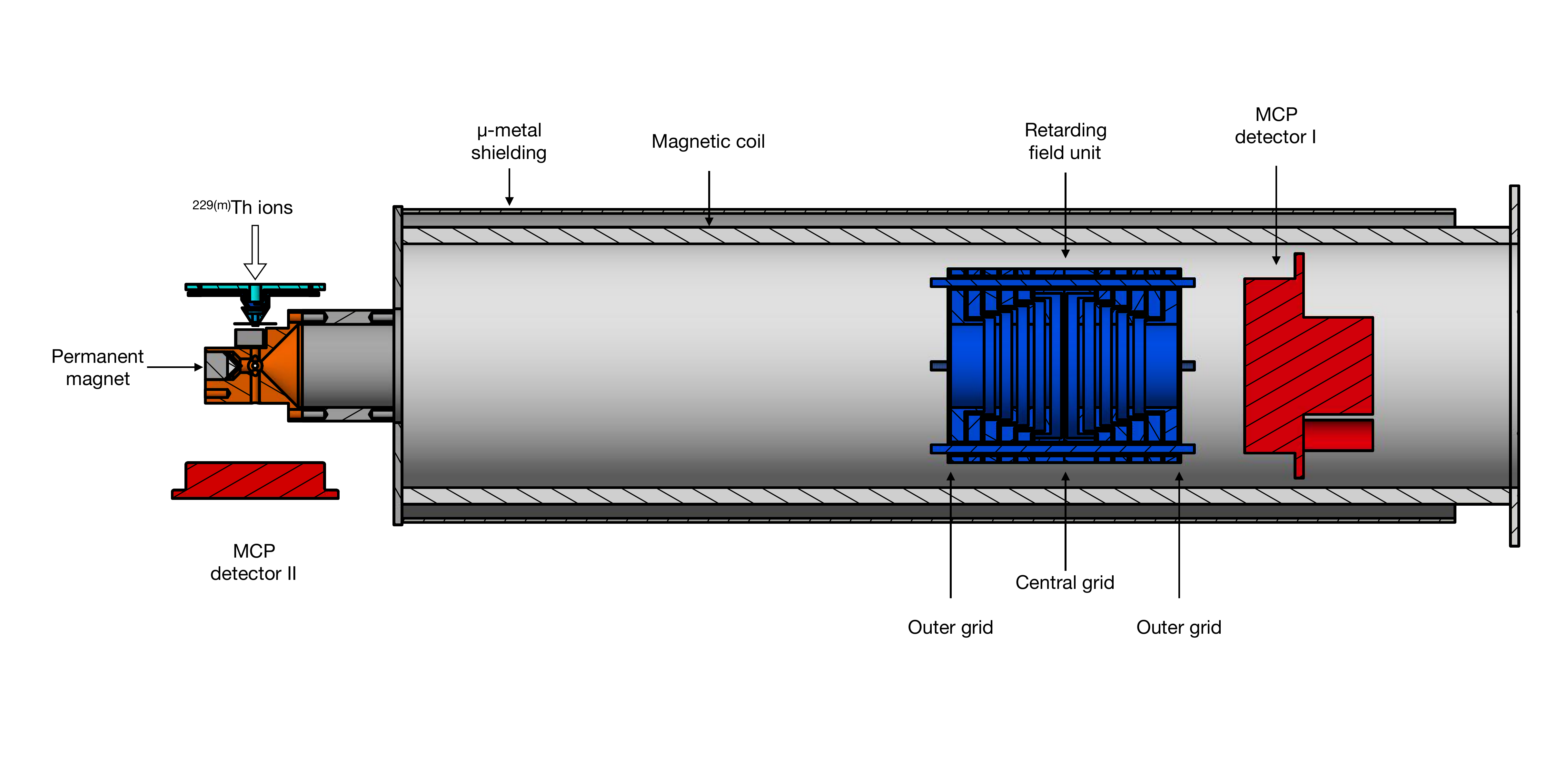}
\caption{Sectional view of the magnetic bottle-type retarding field electron spectrometer.
The retarding field voltage is applied to the central grid in the retarding field unit.
The outer grids are kept on ground, the electrodes in between are biased via a voltage divider chain. \label{Spec}}
\end{figure*}

\noindent\textbf{Experimental setup.} 
The experimental setup consists of three parts: the formation of a $^{229\mbox{\scriptsize m}}$Th ion beam from a $^{233}$U $\alpha$-recoil source, the neutralisation of these ions, and the measurement of the kinetic energy of the IC electrons.\\
\noindent\textbf{$^{229\mbox{\scriptsize m}}$Th ion beam formation.} 
The setup for the generation of an isotopically pure $^{229\mbox{\scriptsize m}}$Th ion beam is reported in detail in refs. \citenum{Lars}, \citenum{Seiferle}, \citenum{Wense3} and \citenum{WenseDiss}.
The experimental setup (see Fig. \ref{Fig1}b) consists of a $^{233}$U $\alpha$-recoil source\cite{Eberhardt} (290 kBq,~ \O~=~90 mm), which is placed in a buffer-gas stopping cell filled with 32 mbar ultrapure helium.
$^{229\mbox{\scriptsize (m)}}$Th recoil ions are stopped and thermalised in the buffer-gas and guided with an RF- and DC-funnel towards a de-Laval nozzle that connects the buffer-gas stopping cell to the subsequent differentially pumped vacuum chamber ($5\times10^{-4}$ mbar).
This chamber houses a segmented radio-frequency quadrupole (RFQ) that serves as an ion guide and buncher.
Ion bunches are formed in the next-to-last quadrupole segment, where a trap is created to store and cool the ions. 
After cooling, the ion bunches are injected in a subsequent quadrupole mass separator (QMS) \cite{Haettner}.
Bunches are generated at a 10 Hz repetition rate and contain about 400 $^{229}$Th$^{3+}$ ions (about 200 $^{229}$Th$^{2+}$ ions).
The ions have a kinetic energy of 51 eV and 34 eV, respectively. \\
The QMS allows to select ions with a specific mass-to-charge ratio.
In the presented experiments, however, it is used as an ion guide and Th ions in different charge states are separated by their time of flight:
The distance from the RFQ-buncher to the detection region is $\approx$~400 mm and the  time-of-flight (from the ion bunch release to the detection of the ion bunches) is 80 $\mu$s (97 $\mu$s) for $^{229}$Th$^{3+}$ ($^{229}$Th$^{2+}$) ions.
This allows to perform experiments with $^{229}$Th$^{3+}$ and $^{229}$Th$^{2+}$ ions in parallel.
As the experiment is optimised for $^{229}$Th$^{3+}$ ions,  $^{229}$Th$^{2+}$ ions are only used to check for consistency.\\
\noindent\textbf{Ion neutralisation.}
The generation of a collimated atom beam is critical for a high count rate of IC electrons in the spectrometer.
The $^{229\mbox{\scriptsize (m)}}$Th atoms are generated by sending $^{229\text{(m)}}$Th$^{3+}$ ions through a dual layer of graphene (EM TEC dual layer graphene, Micro to Nano V.O.F.).
At the point of neutralisation the ion beam already needs to be collimated, which is achieved by adjusting four focusing electrodes such that the count rate of the atoms (generated from $^{229}$Th$^{3+}$ ions) measured with an MCP detector placed behind the neutralisation region is maximised.
The graphene layers are attached to a transmission electron microscopy (TEM) grid which is set to $-$300 V and accelerates the ions.
After neutralisation the atoms retain the kinetic energy of the ions right before neutralisation (945 eV for neutralised Th$^{3+}$ ions).
The bending electrodes are set to +100 V and +1800 V. These voltages ensure that secondary electrons that are generated by the ions in the graphene and also ions which were not neutralised cannot enter the spectrometer region. 
The neutralisation of $^{229\text{(m)}}$Th ions in our setup is studied in detail in ref. \citenum{Amersdorffer}.\\
\noindent\textit{Electron Collection Region}:
Special care has been taken in the design of the spectrometer's electron collection region.
Sectional views of the setup are shown in Extended Data Figure \ref{Triode}.
For a good energy resolution, DC electric fields from the deflection electrodes as well as RF electric fields from the QMS need to be shielded.
Therefore, a shielding grid is placed at the entrance of the spectrometer.
In order not to generate secondary electrons from atoms impinging on any surfaces, the central region above the permanent magnet is broadly cut out.
The entrance to the magnetic coil and the retarding field unit are shielded with a \O~$\approx$~3 mm aperture.
The collection region is coated with gold for stable surface potentials.
The distance between the graphene layer and the collection region is 16 mm.
The time-of-flight for Th atoms with kinetic energy of 945 eV (for neutralised Th$^{3+}$ ions) amounts to 570 ns for this distance. \\
%
%
%
\noindent\textbf{Electron spectrometer.}
A detailed description of the spectrometer as well as characterisation measurements can be found in ref. \citenum{Seiferle3}.\\
A magnetic bottle-type retarding field electron spectrometer is employed. 
The spectrometer components were customised and manufactured from non-magnetic components (either aluminum or titanium) if not stated otherwise. 
Drawings are provided on request by the corresponding author.\\
The electron spectrometer makes use of magnetic and electric fields, 
which are detailed in the following.\\
\textit{Magnetic Fields}: 
In a magnetic bottle-type spectrometer electrons are collected in a strong inhomogeneous field and guided by weak homogeneous magnetic fields.
A strong permanent magnet (VACODYM, Vacuumschmelze GmbH. \O~=~10 mm, height = 10 mm) with an iron cone on top (height = 5 mm, 90$^\circ$ opening angle) generates an inhomogeneous magnetic field of strength $\approx$~200 mT.
The homogeneous magnetic field is generated by a coil (with a length of 400 mm and a diameter of 80 mm, 368 windings in total, Kapton insulated copper wire, \O~=~1 mm) which is placed directly in vacuum. 
The coil is surrounded by a 3 mm thick $\mu$-metal cylinder which acts as a shielding against perturbing external magnetic fields.
The typical field strength in the coil is 2 mT (with a current of 2 A applied). 
No cooling is applied to the coil.
In operation its temperature equilibrates at $\approx$~350 K.
The quality of the electron beam collimation in a magnetic bottle spectrometer (and thus its resolution) depends on the ratio of the magnetic field strength in the electron collection region above the permanent magnet and the coil.

\noindent\textit{Electric Fields}:
Electric fields are used to analyse the kinetic energy of the collimated electron beam:
if an electron's kinetic energy is large enough to surpass a certain retarding field, it is counted with a micro-channel plate (MCP) detector. 
The fields are generated by a retarding field unit which consists of 10 ring electrodes.
Voltages are applied with a commercial voltage supply (MHV-4, four channel voltage supply, mesytec GmbH \& Co. KG.) directly to the central grid and via a voltage divider ($R$ = 1.25 M$\Omega$ for each resistor) to the surrounding electrodes, while the outermost electrodes are kept on ground.
The fields in the retarding field analyser are terminated with three gold grids (electroformed gold mesh, MG20, transmission 90 \%, Precision Eforming LLC) as shown in Extended Data Fig. \ref{Spec}.

\begin{figure*}[ht]

\includegraphics[width = 0.95\textwidth]{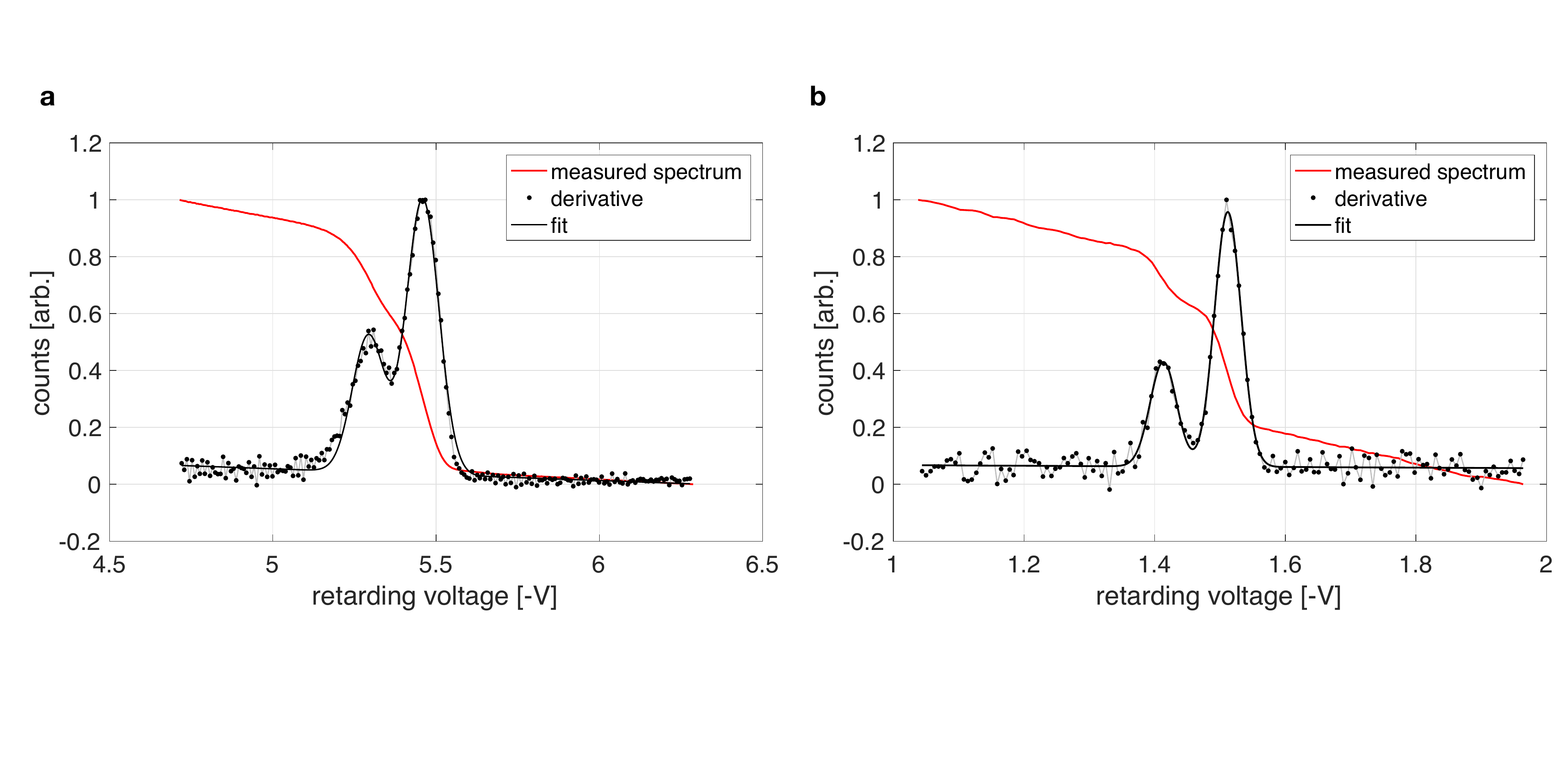}
\caption{Measured (integrated) calibration spectra (red) and their derivatives, together with a fit function (black). For better visualisation, data and derivatives were normalised to their respective maximum. The energy scale has already been corrected for surface potentials.
Measurements were obtained as detailed in ref. \citenum{Seiferle3}.
\textbf{a:} Typical argon calibration spectrum that was used in the measurements (electron kinetic energies 5.46 eV and 5.28 eV). A spectrum recorded with neon, where a clear separation of the two lines (electron kinetic energies 1.52 eV and 1.43 eV) can be achieved is shown in \textbf{b}. 
\label{calibrationMeas}}
\end{figure*}

\begin{figure*}[bt]
\centering
\includegraphics[width = 1\textwidth]{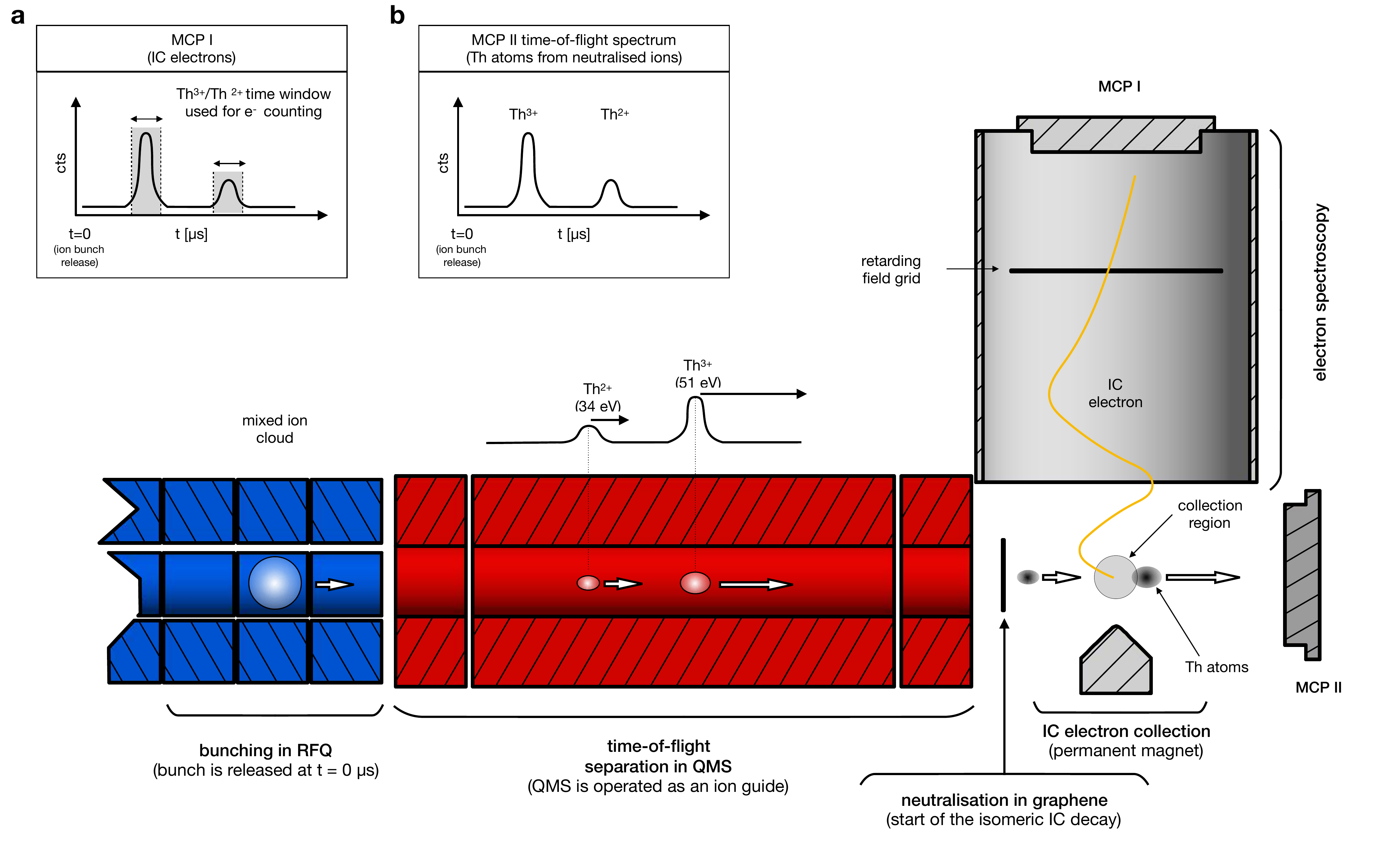}
\caption{Visualisation of the experimental scheme (dimensions are not to scale). A mixed ion cloud (containing Th$^{2+,3+}$ ions) is trapped in a segmented radio frequency quadrupole (RFQ, blue area). The cloud/bunch (white) is released and injected into the QMS (red area) that serves as an ion guide. The difference in kinetic energy, Th$^{3+}$ ions are faster than Th$^{2+}$, leads to a temporal separation of both charge states.
The ion bunches are then neutralised in graphene and continue their flight as atoms (black). 
The atoms are counted with a multi-channel plate (MCP) detector (MCP II). 
The time-of-flight spectrum is shown in inset \textbf{b} (the corresponding measurement is shown in Extended Data Figure \ref{comparison2}b).
The neutralisation also triggered the internal conversion decay of the $^{229\text{m}}$Th nuclear isomer.
When $^{229\text{m}}$Th decays inside the collection region of the electron spectrometer, the IC electron is guided towards MCP I and generates a signal.
The time resolved signal as well as the time windows that are used to obtain the IC electron counts are shown in inset \textbf{a} (a measurement is shown in Extended Data Figure \ref{comparison2}a).
\label{ExperimentalScheme}}
\end{figure*}

\begin{figure*}[bt]
\centering
\includegraphics[width = 0.9\textwidth]{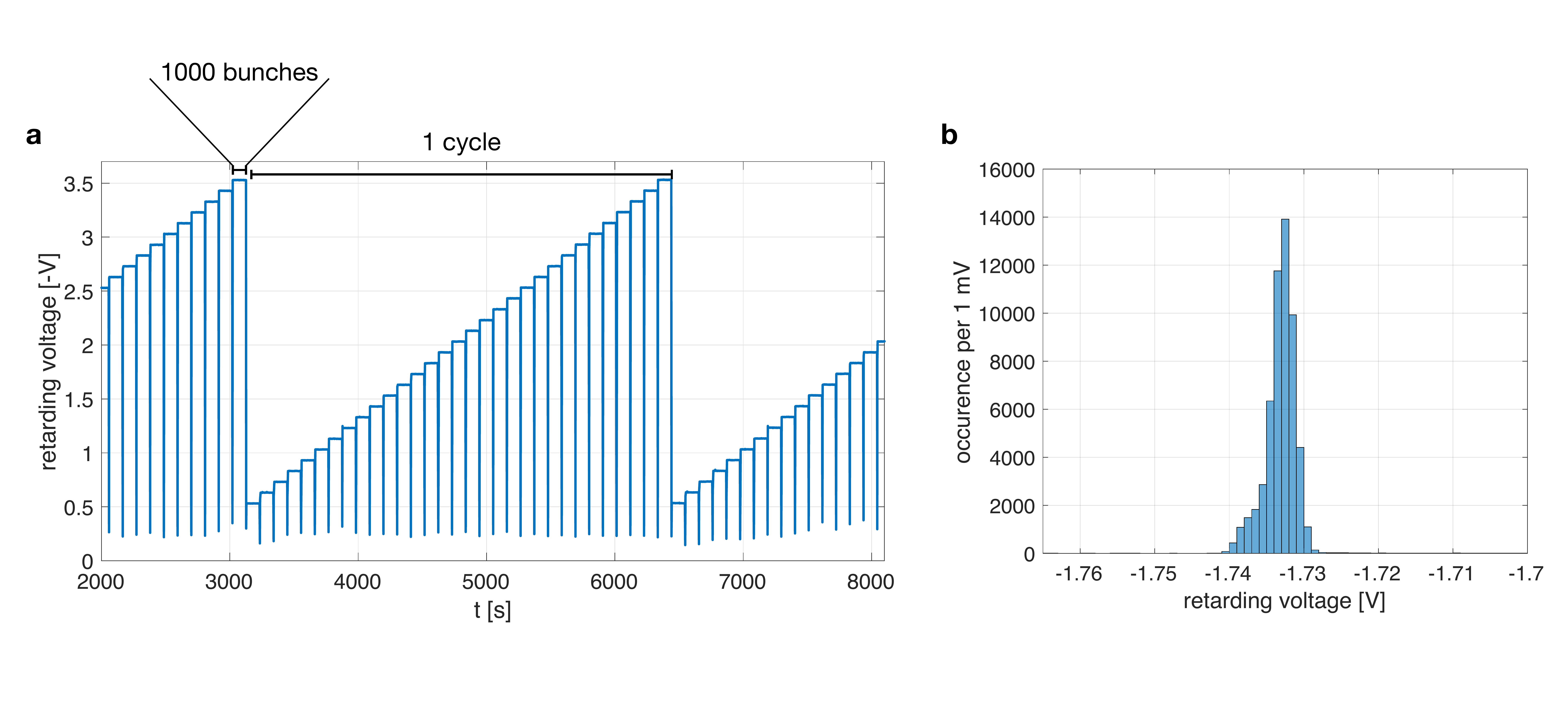}
\caption{Time resolved behaviour of the applied retarding voltages (\textbf{a}).
Retarding voltages between $-0.5$ V and $-3.5$ V have been applied in steps of $-0.1$ V. 
Each retarding voltage is held for 1000 bunches ($\approx$~100 s). 
Before incrementing the voltage it is set to 0 V to make the measurement independent from the order in which the voltages are applied. 
Panel \textbf{b} shows the distribution of one exemplary retarding voltage within 3 days of measurement. 
The stability is better than 10 mV and limited due to temperature fluctuations.  \label{voltagesProjection}}
\end{figure*}

\begin{figure*}
\begin{center}
\includegraphics[scale = 0.25]{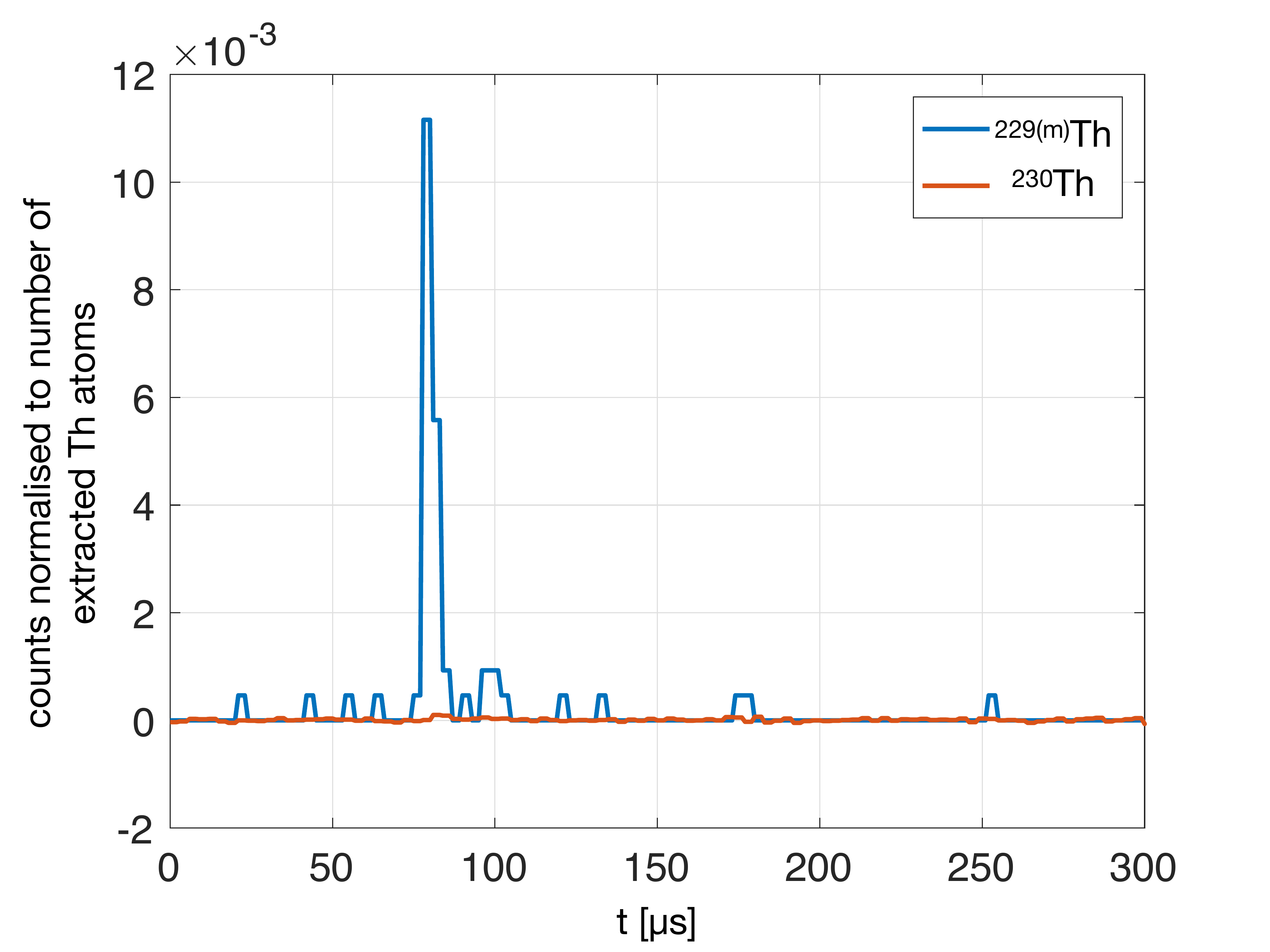}

\end{center}
\caption{Comparative measurement with $^{230}$Th and $^{229\text{(m)}}$Th at 0 V blocking voltage. For the $^{229\text{(m)}}$Th ($^{230}$Th) measurement 5 000 ($>$800 000) bunches were recorded.
In order to compare $^{230}$Th to $^{229\text{(m)}}$Th the counts were normalised to the number of extracted atoms measured with MCP II (see Extended Data Fig. \ref{ExperimentalScheme}). Constant background has been subtracted in the $^{230}$Th measurement. \label{comparison}}
\end{figure*}

%
%
%
%
%
%
%
%
%
%
%
%
\noindent\textbf{Spectrometer performance and calibration.}
The spectrometer is calibrated in order to provide an absolute energy scale.
The energy scale is shifted by the contact potential (\textit{i.e.} the difference of the work function in the collection region and the retarding field analyser), which typically accounts to $-$0.5 eV.
During the operation of the experiment (buffer-gas stopping cell filled, voltages applied to all electrodes) variations of the contact potential are in the order of 0.01 eV over several days.
The stability is good enough not to influence the energy determination. \\
For the calibration and characterisation of the spectrometer, a helium gas-discharge lamp (UVS-20, Specs GmbH.) is used (Extended Data Fig. \ref{Triode}). 
The voltages that are applied to the retarding field electrodes during calibration and the measurement are monitored and recorded under the same conditions with the same voltmeter (34405A, Digital Multimeter 5 1/2-digit, Agilent).
A gaseous calibrant is fed to the spectrometer vacuum cell and photo-electrons are generated along the path of the UV light beam (Extended Data Fig. \ref{Triode}).
A pressure in the lower $10^{-5}$ mbar region generates the required count rates for calibration.
As a regular calibrant argon was used, which emits electrons with an energy of 5.28 eV (Ar$^+ ~^2 P_{1/2}$) and 5.46 eV (Ar$^+ ~^2 P_{3/2}$) \cite{Eland}.
A measurement is shown in Extended Data Figure \ref{calibrationMeas}a.
The linearity of the spectrometer is demonstrated with photoelectrons emitted from molecular nitrogen/air. 
The kinetic energy of the electrons emitted from N$_2$ is in the range of $\approx$~3 eV to 6 eV (ref. \citenum{Harada, Gardner}).
The performance in the energy region which is of interest for the internal conversion electrons can be shown with He I$\beta$ radiation that ionises neon atoms and generates electrons with kinetic energies of 1.43 eV (Ne$^+ ~^2 P_{1/2}$) and 1.52 eV (Ne$^+ ~^2 P_{3/2}$)\cite{Eland}. A corresponding measurement is shown in Extended Data Figure \ref{calibrationMeas}b.
The spectrometer reaches a relative energy resolution of 3\% at full width half maximum. \\
%
%
%
%
%
%
%
%
%
%
\noindent\textbf{Measurement Procedure.}
In the measurements Th ion bunches are sent through graphene where they neutralise.
The experimental scheme is depicted in Extended Data Fig. \ref{ExperimentalScheme}.
Data is taken with two detectors in parallel:
MCP detector I counts the internal conversion (IC) electrons which are emitted from $^{229\text{m}}$Th as the isomer decays inside the spectrometer's collection region.
The IC electron count rate as a function of the applied retarding voltage results in the electron spectrum that is used for the energy determination.
MCP detector II measures the number of extracted $^{229\text{(m)}}$Th atoms.\\
Internal conversion electron spectra are typically recorded for several days.
The applied retarding voltages are permanently logged throughout the whole measurement and the temporal behaviour of the retarding voltages is shown in Extended Data Figure \ref{voltagesProjection}.
In order to be insensitive to possible instabilities in the ion extraction, these measurements consist of a number of cycles. 
One cycle consists of a predefined number of retarding voltages which are held until 1000 bunches have been released.
IC electrons that surpass the applied retarding potential are counted with MCP I.
They are registered in a 6 $\mu$s wide region of interest as shown Extended Data Fig. \ref{ExperimentalScheme}.
The centre of the region of interest coincides with peaks measured in the extraction measurements (performed with MCP II shown in Fig. \ref{Fig1} and Extended Data Fig. \ref{ExperimentalScheme}).
The time-of-flight from the ion bunch release to the  spectrometer/MCP II is measured to 80 $\mu$s.
Detector noise from MCP I is subtracted from the IC electron counts.\\
\textit{Count Rate:} The count rate at $0$ V retarding voltage is in the range between 5 and 10 detected electrons in 1000 ion bunches, which leads to an absolute count rate of 5-10$\times 10^{-2}$ electrons per second.
Assuming 400 ions per bunch, this number can be explained as follows:
2\% of the ions are in the isomeric state (8 ions in the isomeric state per bunch). 
We assume 50 \% of the ions that pass the TEM grid (50\% transmission) are neutralised and reach the spectrometer collection region.
About $\approx$0.7\% of the atoms decay within the spectrometer collection region (see Extended Data Fig. \ref{ExperimentalScheme}):
passing through the graphene foil at a velocity of 2.8$\times 10^4$ m/s, the ions reach the spectrometer collection region after 570 ns. Transit through this 2 mm wide region takes 71 ns. 
Assuming an isomeric lifetime of 10 $\mu$s, 0.7\% of the nuclei decay within the collection region.
The transmission efficiency of the retarding field analyser (3 grids with a transmission of 90\%) is 73\% and the detection efficiency of the MCP detector is 50\%.
In total, this leads to a combined collection, transmission and detection efficiency of $7.5\times10^{-4}$ and yields an expected number of 6 detected electrons per 1000 ion bunches, which is comparable to the actually detected number of IC electrons.\\
We expect a purely Poissonian background of 3 dark counts per second. 
As the time window in which the counts are expected is 6 $\mu$s/bunch, we obtain a signal-to-background ratio of about 150 to 300.
It has been observed that the count rate slowly decreases within several weeks, which is attributed to a degradation of the graphene layers due to the ion bombardment.\\
\noindent
\textbf{Signal Origin.} Comparative measurements with $^{230}$Th ions have been performed under identical conditions as the measurements with $^{229\text{m}}$Th ions and no comparable signal has been measured (see Extended Data Figure \ref{comparison}). 
\noindent Extended Data Figure \ref{comparison} shows the recorded time-of-flight spectra measured at 0 V retarding voltage with $^{229(m)}$Th and $^{230}$Th  ions.
Th$^{3+}$ ions generate a signal at the detector $\approx$~80 $\mu$s  after the bunches have been released.
The signal at $\approx$~97 $\mu$s is attributed to Th$^{2+}$.
The background (mean of the counts between 250 $\mu$s and 500 $\mu$s) has been subtracted and the counts were normalised to the number of extracted atoms which were measured with MCP II (see Extended Data Fig. \ref{ExperimentalScheme}).
For the measurements performed with $^{230}$Th no signal is detected.
This shows clearly that secondary electrons generated in the graphene foil or originating from ionic impact can be fully suppressed and that the detected signal does not correspond to any shell related effects (\textit{e.g.} auto-ionising states).\\
%
%
%
%
%
%
%
%
%
%
\noindent\textbf{Prediction of IC electron spectra.}
The IC process for the case of a magnetic transition is described by the Hamiltonian~\cite{Palffy_NEECTheory_PRA_2006}
\begin{equation}
\hat{H}_{\mathrm{magn}} = - \frac{1}{c} \vec{\alpha} \int d^3r_n \frac{\vec{j}_n(\vec{r}_n)}{|\vec{r}_e-\vec{r}_n|},
\label{hmagn}
\end{equation}
representing the interaction of the electronic and nuclear electromagnetic currents. Here $\vec{j}_n(\vec{r}_n)$ is the nuclear current at the point $\vec{r}_n$, $\vec{r}_e$ is the electronic coordinate, $c$ is the speed of light,   $\vec\alpha$ is the vector consisting of the Dirac matrices $\alpha_i$ for $i=1,\;2,\;3$ and the integration is carried out over the entire nuclear volume. The IC rate is then obtained using Fermi's golden rule involving the matrix element of the operator~(\ref{hmagn}), averaging over the initial nuclear and electronic bound states differing by the magnetic quantum numbers, summation over the final bound states and integration over the momentum directions of the converted electron. 
We consider the Russell-Saunders (LS) coupling scheme for the angular momenta of the four outer electrons taking into account the relative contributions of the LS-components available in ref. \citenum{DBlevels} (the first and the second components for the final states and initial even states and the first component for the initial odd states). We make sure that inclusion of further LS-components does not affect the result by evaluating their contribution with the package GRASP2K\cite{grasp}.

We consider the magnetic dipole ($M1$) channel of IC for the $7s$ electron and neglect contribution of the other orbitals and the electric quadrupole ($E2$) channel based on the analysis presented in ref.~\citenum{Bilous_PRC_2018}. After application of the Wigner-Eckart theorem, the expression for the IC rate takes the form
\begin{equation}\label{IC_rate_GS_M1}
\Gamma_\text{IC}=\frac{8\pi^2}{9}\Lambda B_\downarrow^{M1}\mkern-15mu\sum_{\substack{\kappa\\ \text{ (even }l)}}\mkern-15mu(2j+1)(\kappa-1)^2
\begin{pmatrix}
\frac{1}{2} & j & 1 \\
\frac{1}{2} & -\frac{1}{2} & 0
\end{pmatrix}^2
\left|R_{\varepsilon\kappa}\right|^2,
\end{equation}
where $j$ is the total angular momentum and $\kappa$ is the Dirac angular momentum quantum number of the continuum electron, $B_\downarrow^{M1}$ is the reduced nuclear transition probability and the summation is carried out over $\kappa$ corresponding to even orbital angular momentum number $l$. The electronic radial integral $R_{\varepsilon\kappa}$ is given by
\begin{equation}
R_{\varepsilon\kappa} = \int_{0}^{\infty} \frac{dr}{r^2} \Bigl( P_{n_i \kappa_i}(r)Q_{\varepsilon \kappa}(r) + P_{\varepsilon \kappa}(r) Q_{n_i\kappa_i}(r) \Bigr)\;,\label{IC_radM}
\end{equation}
where $P_{\gamma \kappa}$ and $Q_{\gamma \kappa}$ are the relativistic Dirac radial wave functions. The notation $\gamma$ is the principal quantum number $n$ for bound electron orbitals and the energy $\varepsilon$ for free electron wave-functions, respectively. The total wave function for the electron is expressed via $P_{\gamma \kappa}$ and $Q_{\gamma \kappa}$ as
\begin{equation}\label{IC_radWF_def}
\ket{\gamma \kappa m} = \frac{1}{r}
\begin{pmatrix}
P_{\gamma \kappa}(r) \Omega_{\kappa m}(\hat{r}) \\
iQ_{\gamma \kappa}(r) \Omega_{-\kappa m}(\hat{r})
\end{pmatrix}\;,
\end{equation}
where the functions $\Omega_{\kappa m}(\hat{r})$ of the argument $\hat{r}=\vec{r}/r$ are the spherical spinors. The coefficient $\Lambda$ in~(\ref{IC_rate_GS_M1}) for the case when the IC process expels an electron participating in an internal angular momenta coupling takes the value
\begin{equation}\label{IC_Lambda_expr}
\Lambda=(2J_f+1)(2S_f+1)\left\{
\begin{matrix}
S_i & S_f & \frac{1}{2} \\
J_f & J_i & L_i
\end{matrix}
\right\}^2
\end{equation}
and $\Lambda=1$ otherwise. Here $S$, $J$ and $L$ are the spin, total angular momentum and orbital angular momentum quantum numbers for the initial ($i$) and the final ($f$) electronic configurations, respectively. The IC rate calculation involves the reduced nuclear transition probability for which we considered the theoretical prediction $B_\downarrow^{M1}=0.0076\text{ W.u.}$ ~(ref. \citenum{Minkov}). 
The required relativistic wave functions for the bound electron are evaluated with the GRASP2K package~\cite{grasp} employing the multi-configurational Dirac-Hartree-Fock method. 
The continuum electron wave functions are obtained using the program \textit{xphoto} from the RATIP package~\cite{ratip} for calculations of photo-ionisation cross sections. 

%
%
%
%
%
%
%
%
%
\noindent\textbf{Resonant neutralisation.}
When multiply charged ions (MCI) approach a surface energy levels are shifted due to the interaction with the image charge of the ion and the additional shielding of the ionic core by the target electrons. During the approach of the MCI, the dominant neutralisation process is resonant transfer of electrons from the valence band to (excited) resonant states of the projectile\cite{Burgdorfer}. Upon impact on the surface, loosely bound electrons are stripped from the projectile and quasi-molecular states form in the target system\cite{Arnau}. In the case of very thin target layers, enough electrons will remain with the projectile after separation to form a neutral yet moderately excited atom.
Graphene is known to be an efficient electron donor for MCI passing the layer\cite{Gruber}. We therefore expect (and also observe) that most of our Th$^{3+}$ ions leave the bilayer graphene sheet as neutral atoms.
To understand which orbitals become resonant with the valence band of graphene we have performed density-functional theory (DFT) calculations of the combined Th-graphene system for different distances of the atom from the topmost layer (Ext. Data Fig. 8). 
We employ the VASP software package using the Perdew-Burke-Ernzerhof (PBE)\cite{PBE} exchange-correlation functional\cite{Kresse1, Kresse2}. 
We use the 5.14 PBE projector-augmented wave (PAW) potentials for C and Th provided with VASP, a convergence threshold of  $\Delta$E$<10^{-8}$~eV and a cut-off energy of 400 eV. 
We use a supercell containing 5$\times$5 carbon hexagons in the x-y plane with additional 22 \AA~of vacuum in z direction to avoid any influence of neighboring cells. 
A single Th atom is positioned on a line perpendicular to the graphene sheet and through the center of a graphene hexagon. 
We include spin polarisation and k-point sampling up to 5$\times$5$\times$1 for the supercell. Given the short time-scale of the Th transit, we do not consider atomic relaxation. 
The PAW potentials for carbon include 4 active electrons and a frozen 1s$^2$ core, for thorium 12 active electrons and a frozen [Xe] 4f$^{14}$5d$^{10}$ core, resulting in 412 active electrons in the calculation. 
For each value of the z-coordinate of Th, we extract the Kohn-Sham eigenenergies and calculate the projection of all Kohn-Sham orbitals on spherical harmonics with  $l$= \{s,p,d,f\} centred around the Th atom and a radius of 2 \AA. These calculations indicate that 6s and 6p electrons in Th remain atomic in the graphene target (orange and light-blue symbols at asymptotic energies of $-$40 and $-$20 eV, respectively), while the outermost electrons of thorium have a strong f-characteristic while in the target (purple symbols crossing the valence band of graphene). Orbitals with 7s (orange) and 6d (green) characteristic get into resonance only at larger distances from the surface ($z\approx$ 2~\AA). Upon separation it is thus probable that an electron will remain in an f-state thereby favoring odd excited states for the outgoing Th atom.\\

%
%
%
%
%
%
%
\noindent\textbf{Energy determination.}
As shown above, $^{229\text{(m)}}$Th atoms are in excited states after neutralisation.
Since the exact population of these initial states cannot be measured, we consider all possible initial states consistent with the results of our DFT calculation. 
Comparing the energies of different single-electron orbitals suggests an upper bound of $\approx$20000 cm$^{-1}$ for the excitation energies.
The kinetic energy of the IC electron is evaluated from the energies of the initial ($E_i$) and final ($E_f$) electronic configurations taken from the experimental database~\cite{DBlevels}:
$E_\text{kin} = E_I - (\text{IP} - E_i+E_f)$.\\
The calculated IC rates are shown in Extended Data Fig. \ref{ExpectEven}, where
larger rates are indicated by larger symbols. 
On average, the energy difference $\text{IP} - E_i+E_f$ is larger than the ionisation potential IP (dashed vertical line). 
Furthermore, the energy difference is larger for even initial states than for odd states (blue and red lines in Extended Data Fig. \ref{ExpectEven}, respectively). \\
\noindent The integrated IC electron spectrum generated from one initial atomic excited state $i$ can be represented as a function of the isomeric energy $E_I$ and the applied blocking voltage $U$: $S_i(E_I, U)$.
The spectrum $S_{\mbox{\scriptsize tot}}$ is a linear combination of the spectra
\begin{equation}
S_{\mbox{\scriptsize tot}}(E_I, U) = \sum w_i S_i(E_I, U),
\end{equation}
where $w_i$ are weighting factors that correspond to the population of the initial excited states.
Although the exact population of initial states is unknown, the isomeric energy can be determined by fitting a function of the form $f(U) = a(1-\mbox{erf}[ (U-E_{\mbox{\scriptsize defl}})/b)])$ to the data, which results in $E_{\mbox{\scriptsize defl}}$=1.77$\pm$0.03 eV and $b =$ 0.63 $\pm$ 0.1 eV. This function does not necessarily give the best fit, but it produces a measure which can be used for energy determination.
The energy of the isomer depends on the deflection point via
\begin{equation}
E_I = E_{\text{defl}} + E_0,
\end{equation} \\
where $E_0$ represents a reference energy that is determined by the simulations.
\noindent The analysis procedure is examined with simulated data that samples a random distribution of initial excited states. For a fixed isomeric energy each weighting factor $w_i$ is randomised and the linear combination $S_{\mbox{\scriptsize tot}}$ is calculated. 
The assumed relative energy resolution of the spectrometer is 3\%.
In a next step, the complementary error function is fitted to the simulated data ($f(U) = a (1-\mbox{erf}(U-E_{\mbox{\scriptsize defl}})/b)$).
The deflection point ($E_{\mbox{\scriptsize defl}}$) of the complementary error function shifts according to the simulated isomeric energy, but is only weakly influenced by the population of initial states.
The above steps are repeated several thousand times at a fixed isomeric energy and the position of the deflection point is filled in a histogram if the width $b$ of the error-function is within the interval 0.60$\pm$0.15 eV and thus the shape of the curve is comparable to the measurement.
A Gaussian function can be fitted to the $E_{\text{defl}}$ histogram data (see Fig. \ref{effectiveBinding}c).
The centroid of the generated Gaussian shifts with the isomeric energy and allows to calculate $E_0$.
The reference energy is found to be $E_0$ = 6.506$\pm$0.001 eV, by subtracting the centroid of the $E_{\text{defl}}$ histogram (see Extended Data Table \ref{TabUncert}, 5 states populated, marked with $\ast$) from the simulated energy.
The width of this Gaussian determines the uncertainty which is introduced by the analysis, that will be quantified in the following. 

\noindent
\textit{Quantitative Uncertainty Analysis}: As there are no distinct lines visible in the measured spectrum, it can be excluded that there is only one initial atomic excited state contributing to the measured spectrum.\\
As there are no selection rules involved in the resonant neutralisation process, no excited state can be excluded to contribute to the resulting spectrum and it must rather be expected that every excited state (see Extended Data Table \ref{Table_AllStates}) makes a contribution to the IC electron spectrum. \\

In order to quantify the influence of specific initial state configurations the above analysis has been applied to a sample of 5,000 arbitrary configurations of initial excited states.
It has been found that the number of samples that are used does not have a large influence on the results.
It is found that the standard deviation of the generated histogram (see Fig. \ref{effectiveBinding}c) is mostly influenced by the number of initially excited states that make a dominant contribution to the spectrum. 
Different numbers of initial excited states have been sampled. 
For each calculation a random configuration of $n$ states out of all possible initial states (see Table \ref{Table_AllStates}) is populated. 
The results for an assumed energy of 8.3 eV are listed in Extended Data Table \ref{TabUncert}.
It can be clearly shown, that the distribution narrows when more initial excited states are taken into account. 
\\
\noindent Between 1~eV and 2~eV, the energy resolution of the spectrometer is in the range of 0.03 eV and 0.06 eV (FWHM).
Each initial state in the spectra shown in Ext. Data Fig. \ref{ExpectEven} contributes with 3 to 4 lines to this kinetic energy region.
Assuming 4 inital states being populated there should be up to 16 lines present in this energy region.
These lines should be separated by 0.06 eV.
As it was not possible to resolve any structure in this energy region (also in measurements with a much narrower step size of 0.04 eV) it is concluded, that at least 5 initial excited states contribute to the measured spectrum.   
This case is taken as a conservative scenario which is expected to cover realistic cases within its uncertainty, given by 0.16 eV.
The resulting reference energy is $E_0 = 6.51 \pm 0.16$~eV.

\textit{Energy Determination:}
The energy of the isomeric state is determined by fitting the same error-function as above to the measured data.
The deflection point is found to lie at $E_{\text{defl}}$ = 1.77$\pm$0.03 eV for IC electron spectra. 
Consistency measurements with $^{229\text{m}}$Th$^{2+}$ ions (10-times less IC electrons detected) agree within the statistical uncertainty: 1.73$\pm$0.05 eV.
For the energy determination the value obtained with the $^{229\text{m}}$Th$^{3+}$ ions is used.
In order to relate the position of the deflection point to an isomeric energy, the reference energy needs to be added.
This finally leads to the energy of the isomer:
\begin{equation}
E = 8.28 \pm 0.03_{\mbox{ stat.}} \pm 0.16_{\mbox{ syst.}} \mbox{ eV.}
\end{equation}
\noindent Density functional theory simulations suggest, that among all electronic excited states, initial states with configurations that contain f-orbitals have a higher probability of being populated during the neutralisation process.
Simulations that take these findings into account (by assuming they are populated 10 times stronger than every other state) result in a reference energy of $E_0 = 6.40 \pm 0.03$~eV which is covered by the uncertainty of the scenario described above.
In this case the isomeric energy lies at 8.17$\pm$0.05~eV, corresponding to a wavelength of 151.8$\pm$1.0~nm.\\
\noindent
\begin{table*}\begin{center}
\begin{tabular}{l | c | c}
Number of excited states ($n$) 	&$E_0$  & Width of the distribution $\sigma$\\ \hline\hline
2				& 	6.56$\pm$0.02		&0.22 \\
4				& 	6.517$\pm$0.003		&0.18 \\
5 $\ast$	& 	6.506$\pm$0.001 		&0.16 \\
10				&	6.504$\pm$0.001		&0.12 		\\
20			&	6.492$\pm$0.001		&0.09 \\
30			&	6.494$\pm$0.001		&0.06\\
40			&	6.493$\pm$0.001		&0.05\\ \hline

\end{tabular} 
\caption{Influence of different numbers of initially excited electronic states on the systematic uncertainty of the data analysis. 
While the centroid of the distribution ($E_0$) stays almost constant, the systematic uncertainty ($\sigma$) gets narrower with increasing $n$.
The calculations were performed with an assumed energy of 8.3 eV. 
\label{TabUncert}}
\end{center}
\end{table*}

\begin{table*}
\begin{center}
\begin{tabular}{l r | l r }
\multicolumn{2}{c|}{\textbf{even states}} & \multicolumn{2}{c}{\textbf{odd states}} \\ %
State index & $E_i$ [cm$^{-1}$]	& State index & $E_i$ [cm$^{-1}$]	\\ \hline
1	&	0	&	1	&	7795.275	\\
2	&	2558.057	&	2	&	8243.601	\\
3	&	2869.259	&	3	&	10414.136	\\
4	&	3687.987	&	4	&	10526.544	\\
5	&	3865.475	&	5	&	10783.154	\\
6	&	4961.659	&	6	&	11197.031	\\
7	&	5563.142	&	7	&	11241.730	\\
8	&	6362.396	&	8	&	11877.839	\\
9	&	7280.124	&	9	&	12114.366	\\
10	&	7502.288	&	10	&	13175.113	\\
11	&	8111.005	&	11	&	13945.307	\\
12	&	8800.251	&	12	&	14032.085	\\
13	&	9804.807	&	13	&	14206.917	\\
14	&	11601.031	&	14	&	14243.993	\\
15	&	11802.934	&	15	&	14247.307	\\
16	&	12847.971	&	16	&	14465.222	\\
17	&	13088.563	&	17	&	14481.869	\\
18	&	13297.434	&	18	&	15166.901	\\
19	&	13847.771	&	19	&	15490.077	\\
20	&	13962.522	&	20	&	15618.984	\\
21	&	14204.264	&	21	&	15736.969	\\
22	&	14226.822	&	22	&	16217.482	\\
23	&	15493.221	&	23	&	16346.651	\\
24	&	15863.891	&	24	&	16783.847	\\
25	&	15970.095	&	25	&	17224.303	\\
26	&	16351.943	&	26	&	17354.639	\\
27	&	16554.245	&	27	&	17411.224	\\
28	&	17073.811	&	28	&	17501.176	\\
29	&	17166.108	&	29	&	17847.077	\\
30	&	17398.398	&	30	&	18011.380	\\
31	&	17959.898	&	31	&	18053.617	\\
32	&	18431.686	&	32	&	18069.065	\\
33	&	18549.405	&	33	&	18382.826	\\
34	&	18574.608	&	34	&	18614.338	\\
35	&	18699.623	&	35	&	18809.887	\\
36	&	19273.279	&	36	&	18930.293	\\
37	&	19532.419	&	37	&	19039.153	\\
38	&	19713.031	&	38	&	19227.336	\\
39	&	19832.116	&	39	&	19503.144	\\
	&		&	40	&	19516.981	\\
	&		&	41	&	19588.362	\\
	&		&	42	&	19817.182	\\
	&		&	43	&	19948.395	\\
	&		&	44	&	19986.166	\\
\end{tabular}
\caption{List of all states in the Th atom below 20000 cm$^{-1}$ that are considered in the analysis.
Full data can be found in ref. \citenum{DBlevels}. 
 \label{Table_AllStates}}
\end{center}
\end{table*}
\begin{small}
   
\bibliographystyle{nature}

\end{small}
%
%
%
%
%
%
%
%
%
%
%
%
%
%
%

\begin{figure*}
\begin{center}
\includegraphics[scale = 0.5]{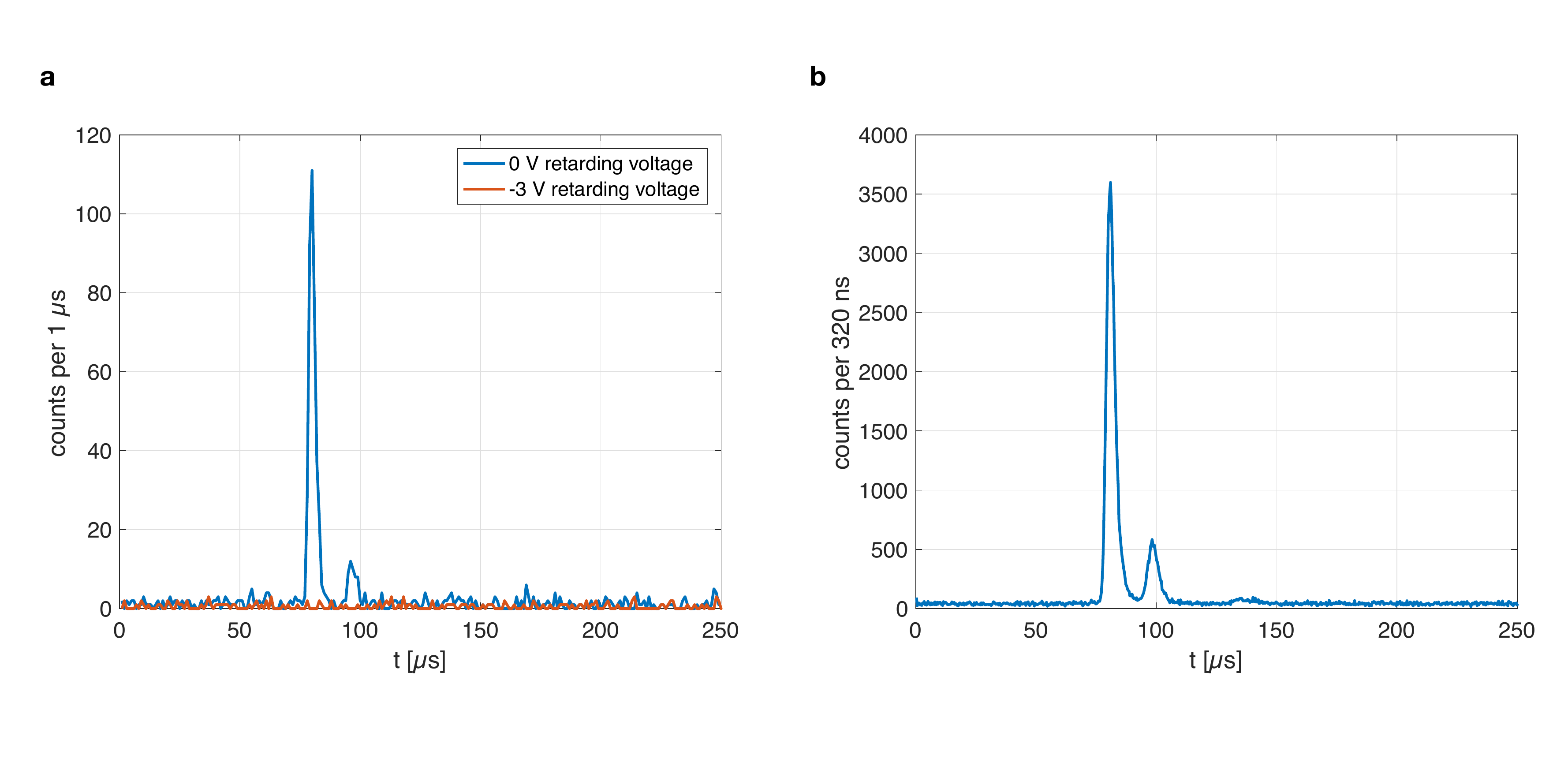}

\end{center}
\caption{Measured signal from the IC electrons (obtained with MCP I) at two different retarding voltages (shown in \textbf{a}). Th$^{3+}$ (Th$^{2+}$) ions generate a signal at t$\approx$~80 $\mu$s (t$\approx$~97 $\mu$s). $10^5$ bunches were recorded for each retarding voltage. Counts from neutralised atoms measured with MCP II are shown in \textbf{b} (where $2.25\times 10^6$ bunches were recorded). 
 \label{comparison2}}
\end{figure*}


\begin{figure*}
\begin{center}
\includegraphics[scale = 0.35]{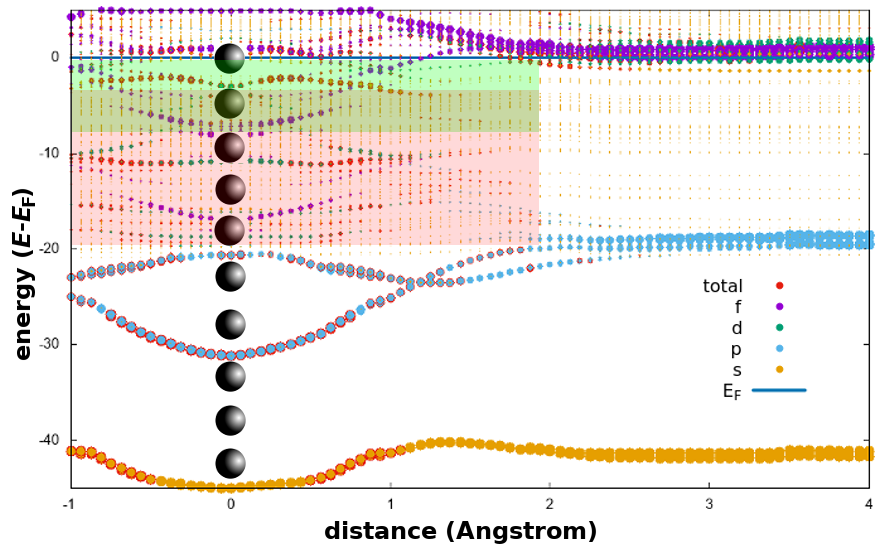}
\end{center}
\caption{Single-electron (Kohn-Sham) orbital energies for a thorium atom as a function of the distance from the topmost carbon layer located at  $z = 0$ \AA. Screening of the core by the target's electron density leads to distance-dependent shifts of the orbital energies. 6s and 6p states (asymptotic energies $\approx$ -40 eV and $\approx$ -20 eV, respectively) remain atomic in graphene, higher orbitals (7s, 6d, 5f) become resonant with the graphene valence bands (overlapping red and green shaded areas indicating the region of high target-electron density) at different distances. In the target, a prevalence of states with f-characteristic (purple symbols) is observed. \label{resNeutral}}

\end{figure*}


\begin{figure*}
\begin{center}
\includegraphics[scale = 0.5]{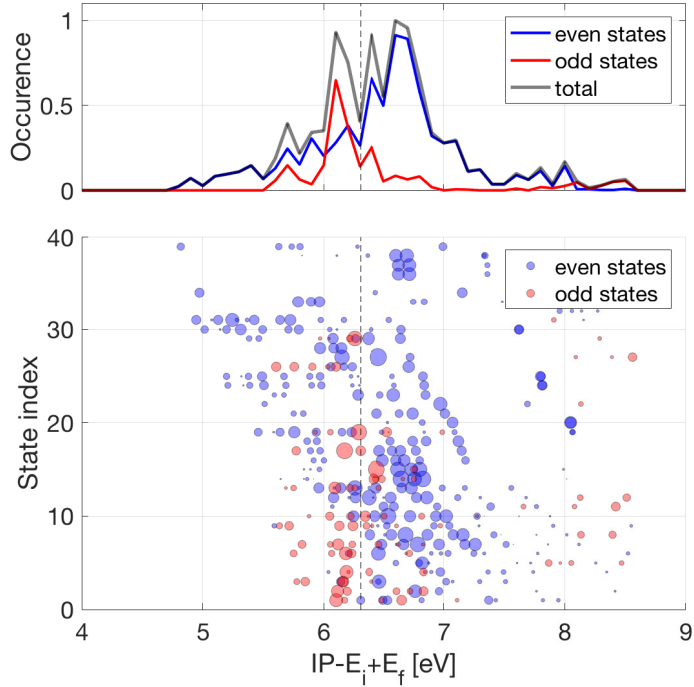}
\end{center}
\caption{Lower panel: Possible values for IP-$E_i$+$E_f$ given by the IC selection rules  for even (blue) and odd (red) initial electronic states of Th. Initial states are numbered according to their state index (given in Extended Data Table 2). The ionisation potential of the Th-atom is shown by the dashed black vertical line. The size of the symbols indicates the rates for any specific transition from an initial to a final state. 
Upper panel: Projection of the lower panel onto the x-axis (energy bin: 0.1 eV). 
 \label{ExpectEven}}
\end{figure*}

\end{spacing}{1.5}

\end{document}